\documentclass{article}[11]

\usepackage{graphics,latexsym}
\usepackage{graphicx}
\usepackage{amsmath, amssymb,amsthm}

\begin{document}

\title{\bf Exact Solutions on Twisted Rings for the 3D Navier-Stokes Equations}

\author{Daniele Funaro}

\maketitle

\centerline{Department of Mathematics} \centerline{University of
Modena and Reggio Emilia}
\centerline{Via Campi 213/B, 41125
Modena (Italy)} \centerline{daniele.funaro@unimore.it}

\begin{abstract}
The problem of describing the behavior of the solutions to the
Navier-Stokes equations in three space dimensions has always been
borderline. From one side, due to the viscosity term, smooth data
seem to produce solutions with an everlasting regular behavior. On
the other hand, the lack of a convincing theoretical analysis
suggests the existence of possible counterexamples. In particular,
one cannot exclude the blowing up of solutions in finite time even
in presence of smooth data. Here we give examples of explicit
solutions of the non-homogeneous equations. These are defined on a
Hill's type vortex where the flow is rotating and swirling at the
same time, inducing the flux to spiraling at a central node.
Despite the appearance, the solution still remains very regular at
the agglomeration point. The analysis may lead to a better
understanding of the subtle problem of characterizing the solution
space of the 3D Navier-Stokes equations. For instance, this result
makes more narrow the path to the search of counterexamples built
on the stretching and twisting of vortex tubes.

\end{abstract}

\vspace{.8cm}
\par\smallskip


\renewcommand{\theequation}{\thesection.\arabic{equation}}

\par\medskip
\setcounter{equation}{0}
\section{An instructive set of explicit solutions}

The literature on fluid dynamics offers a wide range of exact
solutions, suitable to almost any kind of applications (see for
instance \cite{wang} for a review). Many examples are built starting from
{\sl vector spherical harmonic} functions, which are generally used as
expansion basis for more advanced scenarios (see, e.g.,
\cite{rieutord}). Here we deal with some of these functions, which
have not to be confused with their simpler version
related to the {\sl scalar} Laplace equation in spherical coordinates
(see for instance \cite{batchelor}, p.121). For this reason, this section is going
to be a bit technical.  We
discuss a specific situation from the fluid dynamics viewpoint, in
order to prepare the ground to generalizations that will allow in
section 3 to construct families of solutions defined on evolving
vortex rings.

\par\smallskip

First of all, we introduce the family of Bessel functions of the
first kind:
\begin{equation}\label{eq:bes}
J_\alpha^{\prime\prime}(r)~+~\frac{J_\alpha^\prime
(r)}{r}~-~\alpha^2 \frac{J_\alpha (r)}{r^2}~=~-J_\alpha (r)~~~~~r\geq 0
\end{equation}
where, in our context, $\alpha$ is a real positive parameter. It
is known that, for $r$ tending to zero, the following asymptotic
estimate holds:
\begin{equation}\label{eq:as1}
J_\alpha (r)~\approx~\frac{1}{\Gamma (\alpha
+1)}\left(\frac{r}{2}\right)^\alpha
\end{equation}
where $\Gamma$ is the Gamma function. Moreover, $J_\alpha $
turns out to be bounded, since it decays to zero as $\sqrt{2/ \pi
r}$, for $r\rightarrow +\infty$, with an oscillating behavior.
For other properties on Bessel functions the reader is addressed
to \cite{watson}.
\par\smallskip

Successively, it is necessary to introduce the set of Legendre
polynomials:
\begin{equation}\label{eq:leg}
(1-x^2)P^{\prime\prime}_n(x)~-~2x P_n^\prime(x)
~+~n(n+1)P_n(x)~=~0~~~~~ x\in [-1,1]~~n\geq 0
\end{equation}
By the substitution $x=\cos\theta$, the above equation
takes the form:
\begin{equation}\label{eq:let}
\frac{1}{\sin\theta}\frac{\partial}{\partial\theta}
\Big(\sin^2\hspace{-.1cm}\theta
 ~P^{\prime}_n(\cos\theta) \Big)~=~n(n+1)P_n(\cos\theta)
\end{equation}
\par\smallskip

We are now ready to work in spherical coordinates:
\begin{equation}\label{eq:coor}
(x,~y,~z)~=~(r\sin\theta \cos\phi, ~r\sin \theta\sin\phi,~ r
\cos\theta)
\end{equation}
with  $0\leq \phi <2\pi$, $0\leq \theta\leq \pi$ and $r\geq 0$.
\par\smallskip

Let us start by fixing $n\geq 1$ and defining the stationary
vector field $\hat{\bf w}=(\hat w_1, \hat w_2, \hat w_3)$ whose
components are:
$$\hat w_1~=~n(n+1)\frac{J_{n+1/2}(r)}{r\sqrt{r}}~P_n(\cos \theta )$$
$$\hat w_2~=~-\frac{1}{\sqrt{r}}\left( J_{n+1/2}^\prime (r)
~+~\frac{J_{n+1/2}(r)}{2r}\right )\sin \theta ~P_n^\prime(\cos \theta )$$
\begin{equation}\label{eq:vec}
\hat w_3~=~0
\end{equation}
Note that there is no dependence on the variable $\phi$. Moreover,
there is cylindrical symmetry along the $z$-axis. Later, in
section 4, we will generalize this setting.

\par\smallskip
Afterwards, one defines the time-dependent field
$\hat{\bf v}=(\hat v_1, \hat v_2, \hat v_3)$ given by:
\begin{equation}\label{eq:vev}
\hat {\bf v}~=~e^{-\nu t}\hat {\bf w} ~~~~~~~~~~ t\geq 0
\end{equation}
The constant $\nu >0$ will play the role of viscosity parameter.
\par\smallskip

We start by discussing the regularity of $\hat{\bf w}$. Of course,
the same arguments will apply to  $\hat {\bf v}$, $\forall t\geq
0$. It should be clear that $\hat {\bf w}\in C^\infty ({\bf
R}^3-\{ 0\})$, $\forall n\geq 1$. Less evident is the situation at
the origin ($r=0$). For $n\geq 2$, thanks to (\ref{eq:as1}) we
deduce that $\hat{\bf w}$ tends to zero. Therefore, a continuous
prolongation is obtained by setting $\hat{\bf w}=0$ at the origin.
The differentiability follows from an analysis similar to that of
the more delicate case $n=1$, to which we are now going to
concentrate our attention.
\par\smallskip

Let us first note that $P_1(\cos \theta )=\cos \theta~$ and $~P_1^\prime (\cos
\theta )=1$. Moreover, it is known that:
\begin{equation}\label{eq:exj}
J_{3/2}(r)~=~\sqrt{\frac{2}{\pi}}~\frac{\sin r -r\cos
r}{r\sqrt{r}}
\end{equation}
Thus, in a neighborhood of the origin one gets:
\begin{equation}\label{eq:exj2}
n(n+1)\frac{J_{3/2}(r)}{r\sqrt{r}}~\approx~\frac{2}{3}\sqrt{\frac{2}{\pi}}
\left(1-\frac{r^2}{10}+\frac{r^4}{280}\right)
\end{equation}
The second component of $\hat{\bf w}$  follows a similar
asymptotic behavior:
\begin{equation}\label{eq:exj3}
\frac{1}{\sqrt{r}}\left( J_{3/2}^\prime (r)
~+~\frac{J_{3/2}(r)}{2r}\right
)~\approx~\frac{2}{3}\sqrt{\frac{2}{\pi}}
\left(1-\frac{r^2}{5}+\frac{3r^4}{280}\right)
\end{equation}
The important fact is that the coefficients in
(\ref{eq:exj2}) and (\ref{eq:exj3}) not depending on $r$ are the same.
\par\smallskip

In order to understand what is actually happening for $r=0$ it is
better to argue in Cartesian coordinates. The local orthogonal
triplet of the spherical reference framework, has the expression:
$${\bf r}~=~(\sin\theta\cos\phi , ~\sin\theta\sin\phi,
~\cos\theta)$$
$${\boldsymbol{\theta}}~=~(\cos\theta\cos\phi , ~\cos\theta\sin\phi,
~-\sin\theta)$$
\begin{equation}\label{eq:ort}
{\boldsymbol{\phi}}~=~(-\sin\phi , ~\cos\phi, ~0)
\end{equation}
with
$$\sin\theta =\sqrt{x^2+y^2}/\sqrt{x^2+y^2+z^2}~~~~~~~~\cos\theta
=z/\sqrt{x^2+y^2+z^2}$$
\begin{equation}\label{eq:ort2}
\sin\phi =y/\sqrt{x^2+y^2}~~~~~~~~~~~~\cos\phi
=x/\sqrt{x^2+y^2}
\end{equation}
Neglecting higher order infinitesimal terms, near the origin one
has from (\ref{eq:exj2}) and (\ref{eq:exj3}):
\begin{equation}\label{eq:ort3}
\hat{\bf w}~=~\hat w_1{\bf r}+\hat w_2{\boldsymbol{\theta}} +\hat
w_3{\boldsymbol{\phi}}~
\approx~\frac{2}{3}\sqrt{\frac{2}{\pi}}\left(\frac{zx}{10},
~\frac{zy}{10},~1-\frac{z^2}{10}-\frac{x^2+y^2}{5}\right)
\end{equation}
By placing at the origin the vector $(0,0,2\sqrt{2}/3\sqrt{\pi})$
(written in Cartesian coordinates) one gets a continuous vector
field. A more careful analysis shows that $\hat{\bf w}$ can be
differentiated infinite times. A plot of $\hat {\bf w}$ near the
origin of the plane $\phi =0$ is shown in figure 1. Due to the
decay at infinity of  Bessel functions, the whole field is bounded
in ${\bf R}^3$.
\par\smallskip


\begin{center}
\begin{figure}[!h]
\centerline{\includegraphics[width=10.cm,height=8.cm]{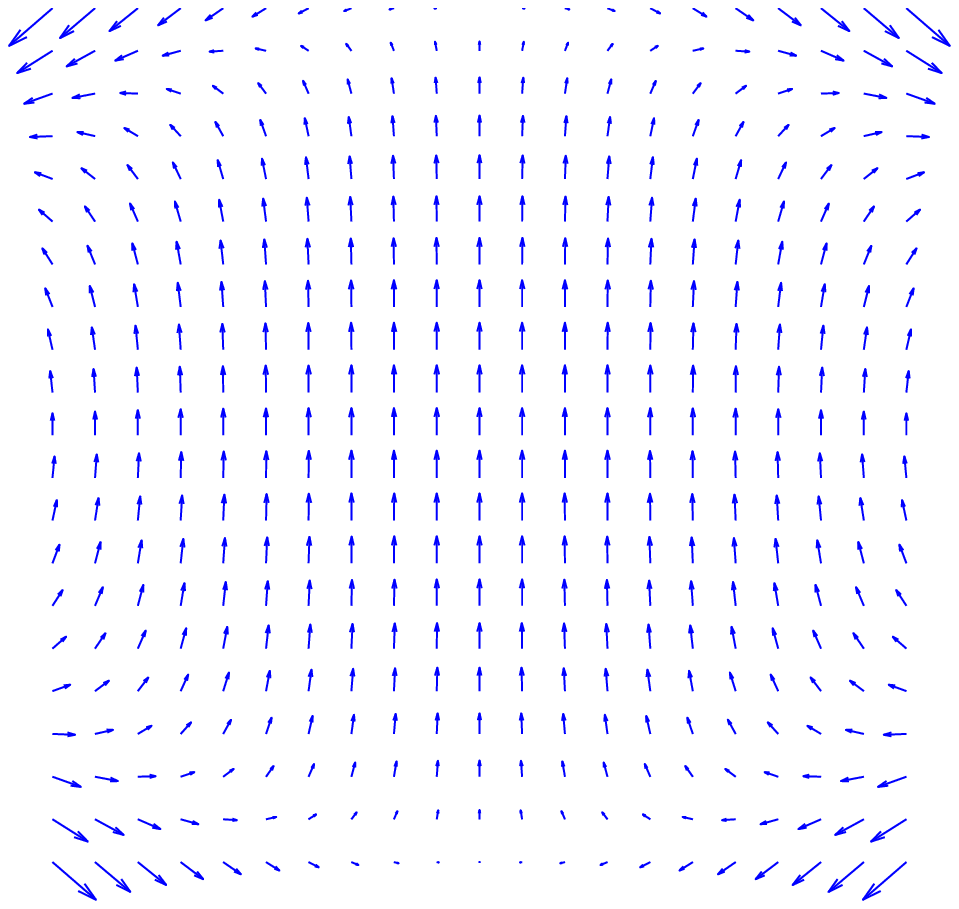}}
\vspace{-.3cm}
\begin{caption}{\small Qualitative displacement of the field $\hat {\bf w}$
near the origin for $n=1$. The plot shows the projection on the
plane $(x,0,z)$ obtained from (\ref{eq:coor}) by setting $\phi=0$.}
\end{caption}
\end{figure}
\end{center}

The vector fields previously introduced satisfy a series of striking
results that we are going to recall. The first important relation
is that, independently of $n$,  $\hat {\bf v}$ has zero
divergence:
\begin{equation}\label{eq:div}
\rm{div}\hat {\bf v}~=~0 ~~~~~ {\rm in}~{\bf R}^3~~~~\forall
~t\geq 0
\end{equation}
The proof is obtained by a direct check. Indeed, for $\alpha
=n+1/2$, thanks to (\ref{eq:let}) one recovers $\forall t\geq 0$
and $r>0$:
$${\rm div}\hat {\bf v}~=~\frac{1}{r^2}\frac{\partial}{\partial
r}(r^2\hat v_1)~+~\frac{1}{r\sin\theta}
\frac{\partial}{\partial\theta}(\sin\theta ~\hat v_2)$$
$$~=~e^{-\nu t}\left[ \frac{n(n+1)}{r^2}\frac{\partial}{\partial r} (\sqrt{r}
J_\alpha )~P_n(\cos\theta )\right.~~~~~~~~~~~~~~~~~~~~~~~~~~~~~~$$
\begin{equation}\label{eq:co1}
~~~~~~~~~~~~~~~~~~~-\left. \frac{1}{r\sqrt{r}\sin\theta}\left(
J^\prime_\alpha
+\frac{J_\alpha}{2r}\right)\frac{\partial}{\partial\theta}
\Big(\sin^2\hspace{-.1cm}\theta
 ~P^{\prime}_n(\cos\theta) \Big)\right] =~0
\end{equation}
Equation (\ref{eq:div}) is also true at the origin. To check this,
one can use either (\ref{eq:ort3}) or symmetry arguments (from the
physical viewpoint in any neighborhood of the origin the inbound
flow equates that directed outbound).
\par\smallskip

Another important fact is that $\hat {\bf v}$ satisfies the
parabolic equation:
\begin{equation}\label{eq:aut}
\frac{\partial \hat{\bf v}}{\partial t}~-~\nu \Delta \hat{\bf
v}~=~0 ~~~~~ ~~\forall ~t\geq 0
\end{equation}
The statement follows from writing the Laplace operator in
spherical coordinates. Considering that there is no dependence on
the variable $\phi$, the expression of the operator looks rather
simplified:
$$(\Delta \hat{\bf v})_1~=~\frac{1}{r^2}\left[\frac{\partial}{\partial r}\left(
r^2\frac{\partial \hat v_1}{\partial r}\right)
~+~\frac{1}{\sin\theta}\frac{\partial}{\partial \theta}\left(
\sin\theta \frac{\partial \hat v_1}{\partial \theta}\right)\right.
~~~~~~$$
\begin{equation}\label{eq:la1}
~~~~~~~~~~\left.~- 2 \left( \hat v_1~+~ \frac{\partial \hat
v_2}{\partial \theta}~ +~\hat
v_2\frac{\cos\theta}{\sin\theta}\right)\right]~=~\frac{1}{\nu}\frac{\partial
\hat v_1}{\partial t}
\end{equation}

$$(\Delta \hat{\bf v})_2~=~\frac{1}{r^2}\left[\frac{\partial}{\partial r}\left(
r^2\frac{\partial \hat v_2}{\partial r}\right)
~+~\frac{1}{\sin\theta}\frac{\partial}{\partial \theta}\left(
\sin\theta \frac{\partial \hat v_2}{\partial \theta}\right)\right.
~~~~~~$$
\begin{equation}\label{eq:la2}
~~~~~~~~~~\left. ~+~ 2\frac{\partial \hat v_1}{\partial \theta}~
-~\frac{\hat
v_2}{\sin^2\hspace{-.1cm}\theta}\right]~=~\frac{1}{\nu}\frac{\partial
\hat v_2}{\partial t}
\end{equation}

\begin{equation}\label{eq:la3}
(\Delta \hat{\bf v})_3~=~\frac{1}{\nu}\frac{\partial \hat
v_3}{\partial t}~=~0
\end{equation}
The check of the above formulas, by direct substitution of the
expressions of $\hat v_1$ and $\hat v_2$, requires however a
generous amount of patience. We do not report such details here.
We just recall that the equations (\ref{eq:bes}) and
(\ref{eq:leg}) must be used with profusion. A simpler verification
is obtained by recovering a potential vector ${\bf A}$ such that
$\hat {\bf w}={\rm rot}{\bf A}$ (see section 3) and showing that
${\bf A} ={\rm rot}\hat{\bf w}$. This implies that
$-\Delta\hat{\bf w}= {\rm rot}({\rm rot}\hat{\bf w})=\hat{\bf w}$
(recall that $\hat{\bf w}$ has zero divergence), and this last
relation implies (\ref{eq:aut}).
\par\smallskip

Similar solutions can be found in the 2D framework. For instance,
working in polar coordinates $(r,\theta )$, we define for $n\geq
1$:
\begin{equation}\label{eq:pol}
\hat{\bf w}~=~(\hat w_1, \hat w_2)~=~\Big( n^2\frac{J_n(r)}{r}T_n(\cos\theta ),
~ -J^\prime_n(r)\sin\theta ~T_n^\prime(\cos\theta )\Big)
\end{equation}
where $T_n$ is the $n$-th Chebyshev polynomial. Again, one has
${\rm div}\hat{\bf w}=0$ and $-\Delta\hat{\bf w}= \hat{\bf w}$.
\par\smallskip

After having introduced the velocity field  $\hat {\bf v}$, we
need to explain its role in the framework of the Navier-Stokes
equations. This will be the subject of the next section. The
origin must be regarded as the bottleneck of a generous amount of
fluid entering  for $z<0$ and exiting for
$z>0$. Nevertheless, this is not enough to generate singularities
in the field or its derivatives. Such an observation continues to
be true even by introducing a swirl in the $\phi$ direction, as we
shall see in section 4.

\section{Connections with the Navier-Stokes equations}

A further property of the field $\hat {\bf w}$ is that its
rotational has a very simple expression:
\begin{equation}\label{eq:rot}
{\rm rot}\hat{\bf w}~=~\Big( 0,~0, ~\frac{1}{\sqrt{r}}J_{n+1/2}(r)
\sin \theta ~P_n^\prime(\cos\theta )\Big)
\end{equation}
where the third component has been computed via the formula:
\begin{equation}\label{eq:ro3}
({\rm rot}\hat{\bf w})_3~=~
\frac{1}{r}\left[\frac{\partial}{\partial r}(r \hat w_2)
~-~\frac{\partial \hat w_1}{\partial \theta}\right]
\end{equation}
Due the asymptotic expression (\ref{eq:as1}), the rotational goes
smoothly to zero at the origin as $r^n$. Considering that $\hat
{\bf w}$ is bounded, the field $\hat {\bf w}\times {\rm rot}\hat
{\bf w}$ also goes to zero at the origin.
\par\smallskip

We now set:
$$\hat {\bf f}~=~-\hat {\bf v}\times {\rm rot}\hat {\bf v}~=~
-~e^{-2\nu t}~\hat {\bf w}\times {\rm rot}\hat {\bf w}$$
\begin{equation}\label{eq:set}
\hat
p=-~\textstyle{\frac{1}{2}}\vert \hat {\bf v}\vert^2~=~
-~\textstyle{\frac{1}{2}}e^{-2\nu t}~\vert \hat {\bf w}\vert^2
\end{equation}

By applying well-known notions of differential calculus,
one may write:
\begin{equation}\label{eq:rel}
(\hat {\bf v}\cdot\nabla )\hat {\bf
v}~=~\textstyle{\frac{1}{2}}\nabla\vert \hat {\bf v}\vert^2~-~
\hat {\bf v}\times {\rm rot}\hat {\bf v}~=~-\nabla \hat p~+~\hat{\bf
f}
\end{equation}
Therefore, using (\ref{eq:aut}), we finally have:
\begin{equation}\label{eq:nav}
\frac{\partial \hat {\bf v}}{\partial t}~-~\nu \Delta \hat {\bf
v}~+~ (\hat {\bf v}\cdot\nabla )\hat {\bf v}~=~-\nabla \hat
p~+~\hat{\bf f} ~~~~~\forall ~t\geq 0
\end{equation}
Recalling the divergence-free condition (\ref{eq:as1}), we may
argue that the couple $\hat {\bf v}$, $\hat p$ satisfies the 3D
Navier-Stokes problem with the forcing term $\hat {\bf f}$ and the
initial condition $\hat {\bf v}=\hat {\bf w}$ at time $t=0$. Note
that our pressure is negative, but this is not a problem, since a
scalar potential is defined up to an additive constant and $\vert
\hat{\bf v}\vert$ is bounded in the whole ${\bf R}^3$ space.
\par\smallskip

The choice of $\hat p$ looks physically correct. The quantity
$\frac{1}{2}\rho\vert \hat {\bf v}\vert^2$, where $\rho$ is the
mass density, is related to kinetic energy. In the incompressible
case, which is the one we are studying here, $\rho$ is constant in
${\bf R}^3$ (assume $\rho =1$). The setting is in agreement with
the Bernoulli principle, stating that the quantity
$\frac{1}{2}\rho\vert \hat {\bf v}\vert^2 +\hat p$ is preserved
along the stream-lines. Thus, the balance between  $-\nabla
\hat p$ and the gradient part of $(\hat {\bf v}\cdot\nabla )\hat {\bf v}$ 
only involves energy properties belonging to the fluid,
while the extra-force $\hat {\bf f}$ (not of conservative type in
our case) represents external constraints.
\par\smallskip

Including  potential functions of conservative fields, such as
gravity, in the pressure term is an allowed procedure (in this
fashion, when considering the case of gravitational potentials,
pressure directly turns out to decay with altitude). Effective
forcing terms can be built with fields having the rotational
different from zero. Note for instance that, if (\ref{eq:set}) was
replaced by the setting $\hat {\bf f}=0$ and $(\hat {\bf
v}\cdot\nabla ) \hat {\bf v}=-\nabla \hat p$, we would get an
incompatible situation since the advective term is not the
gradient of any potential.
\par\smallskip

Our solution is defined in ${\bf R}^3$. Of course, a fictitious
boundary-value problem on an open domain $\Omega$ may be obtained
by assigning on  $\partial\Omega$ the restriction of the field
$\hat {\bf v}$, $\forall t>0$. When $\Omega$ is a sphere of radius
$R$ centered at the origin, there are two interesting situations
to be mentioned. If $R$ is a zero of $J_\alpha$ ($\alpha =n+1/2$),
we have that the radial component on $\partial\Omega$ is
identically zero, so the boundary field is tangential. The
boundary datum is purely radial if instead $R$ is a zero of
$J^\prime_\alpha +J_\alpha /2r$.
\par\smallskip

Assuming that $R$ is the first zero of $J_{n+1/2}$ (the most
interesting case is $n=1$, where $R$ is approximately equal to
$4.49$), we can interpret the solution as a stable vortex ring
$\Omega$. Tangential velocity is imposed on the boundary and the
term $\hat {\bf f}$ plays the role of centripetal force keeping
the fluid in stationary circular motion. In truth the domain
$\Omega$ is a degenerate toroid where the central hole is reduced
to a segment of the vertical axis $z$, i.e., we are simulating a
Hill's type vortex (see, e.g., \cite{acheson}, p.175).
\par\smallskip

The same construction can be considered in the 2D case (see
(\ref{eq:pol})). The corresponding $\hat {\bf f}$ is given by
$J_n(r)\sin\theta~ T_n^\prime(\cos\theta ) (-\hat w_2, \hat w_1)$,
which is tending to zero at the origin (see (\ref{eq:as1})) for
any $n\geq 1$.
\par\smallskip

Let us also note that in order to get a time decay at infinity for
the solution proportional to $e^{-\nu t}$, as the dissipative term
prescribes, our forcing field  $\hat {\bf f}$ must go to zero at
least as $e^{-2\nu t}$, i.e. quadratically faster. This
observation indicates that the time smoothness properties of the
right-hand side should enter more formally in the a priori
estimates, if one looks for regularity theorems.
\par\smallskip

In the theoretical analysis of the Navier-Stokes equations, the
gradient terms are usually eliminated by projection into the space
of functions with null divergence. An alternative approach is to
differentiate the momentum equation. By taking the rotational on
both sides, one gets for some right-hand side ${\bf f}$:
\begin{equation}\label{eq:rof}
\frac{\partial\boldsymbol{\omega}}{\partial t}~-~\nu\Delta
\boldsymbol{\omega}~+~({\bf v}\cdot\nabla
)\boldsymbol{\omega}~-~(\boldsymbol{\omega}\cdot\nabla ){\bf
v}~=~{\rm rot}{\bf f}
\end{equation}
where the vorticity $\boldsymbol{\omega}$ automatically satisfies
${\rm div}\boldsymbol{\omega}=0$. In this way we again eliminate
the gradient term $\nabla (\frac{1}{2}\vert {\bf v}\vert^2+p)$.
Note that $\boldsymbol{\omega}$ does not satisfy homogeneous
boundary conditions, making difficult the theoretical analysis of
(\ref{eq:rof}).
\par\smallskip

From the explicit expression of $\hat {\bf w}$, a certain number
of collateral problems can be examined. For instance, one can try
to find ${\bf v}$ and $p$ such that:
$$
\frac{\partial  {\bf v}}{\partial t}~-~\nu \Delta  {\bf v}~+~ (
{\bf v}\cdot\nabla ) {\bf v}~=~-\nabla  p~+~ {\bf f}
~~~~~~~~~~\forall t>0
$$
\begin{equation}\label{eq:pr1}
{\rm div}  {\bf v}~=~0 ~~~\forall t>0~~~~~~~{\rm and}~~{\bf v}={\bf
v}_0=0 ~~{\rm for}~~t=0
\end{equation}
Here  ${\bf f}=-\gamma \hat {\bf w}\times {\rm rot}\hat {\bf w}$,
where $\hat{\bf w}$ is defined in (\ref{eq:vec}) and $\gamma$ is a
given positive function of time with $\gamma (0)=0$. We suppose
that $\gamma$ grows for a while, then it decays fast to zero. Both
${\bf f}$ and ${\bf v}_0$ are defined in ${\bf R}^3$.  The
solution ${\bf v}$ is expected to develop the same vortices as the
field $\hat {\bf v}$ introduced before. This should happen as soon
as $t>0$. Things may however depend on the rate of growth of
$\gamma$. We guess (but this is just a conjecture to be verified
with numerical simulations) that, if $\gamma$ follows a slow
growth at the beginning, viscosity might dominate smoothing out
the solution. Alternatively, if $\gamma$ has a rapid increase, the
rotatory behavior may catch up and persist.
\par\smallskip

A variant of the example above is reformulated by replacing ${\bf
f}$ in (\ref{eq:pr1}) by a uniformly approximating  sequence ${\bf
f}_\tau$ of functions. We can let the parameter $\tau$ depend on
$t$, in such a way that the limit ${\bf f}=- \hat {\bf w}\times
{\rm rot}\hat {\bf w}$ is reached in a finite time $T$. It is
possible to define ${\bf f}_\tau$  in the time interval $]0,T[$ in
order to get solutions representing classical vortex rings (i.e.,
closed vortex tubes shaped like a doughnut) tending to a Hill's
type vortex as $t\rightarrow T$. We study in the next section a
general recipe to generate this kind of sequences.

\section{Solutions on evolving rings}

Let us now introduce a general approach to get families of
solutions defined on rings. We first observe that a vector
potential ${\bf A}$ may be assigned to the field $\hat{\bf w}$ in
(\ref{eq:vec}). This is such that $\hat{\bf w}={\rm rot}{\bf A}$
and explicitly given by ${\bf A}~=~(0,0,A_3)$, where:
\begin{equation}\label{eq:poa}
A_3~=~\frac{1}{\sqrt{r}}J_{n+1/2}(r) \sin \theta
~P_n^\prime(\cos\theta )
\end{equation}
Information on vector potentials can be for instance retrieved in
\cite{gallavotti}, p.53, where connections with magnetic fields
are also mentioned. It turns out that ${\bf A}$ is zero along the
$z$-axis.
\par\smallskip

Comparing with (\ref{eq:rot}), we deduce that ${\bf A}={\rm
rot}\hat{\bf w}$. As a consequence, by observing that $~{\rm
rot}({\rm rot} {\bf A})=-\Delta {\bf A}+\nabla ({\rm div}{\bf
A})=-\Delta {\bf A}$, the following relation holds:
\begin{equation}\label{eq:rep}
-~\Delta {\bf A}~=~{\bf A}
\end{equation}
Therefore ${\bf A}$ is an eigenfunction of the vector Laplacian,
corresponding to the eigenvalue $\lambda =1$. More precisely, the
sphere $\Omega$ centered at the origin of radius $R$ (the first
zero of $J_{n+1/2}$), deprived of the vertical axis,
is such that the smallest eigenvalue of the
operator $-\Delta$ is equal to 1. The homogeneous boundary conditions
associated to such an operator are of Dirichlet type, including in the
boundary also the vertical segment at the intersection of the
sphere and the $z$-axis. They can be however substituted by
tangential type conditions:
\begin{equation}\label{eq:tbc}
{\rm rot}{\bf A}\cdot {\bf n}~=~0
\end{equation}
with ${\bf n}$ denoting the outward normal field on
$\partial\Omega$, that is what we got for the field
$\hat {\bf w}$.
\par\smallskip

In order to examine other similar situations, we work with
cylindrical coordinates $(r, z, \phi)$. We look for vector
potentials of the form ${\bf A}~=~(0,0,A_3)$, where $A_3$ does not
depend on $\phi$. In addition we would like to impose
(\ref{eq:rep}) and (\ref{eq:tbc}). We are then requiring that
$\lambda =1$ must be the first eigenvalue, so restricting a bit
the choice of the domain $\Omega$. In terms of the only unknown
$A_3$, the vector differential problem becomes:
\begin{equation}\label{eq:pep}
-\left(\frac{\partial^2 A_3}{\partial
r^2}~+~\frac{1}{r}\frac{\partial A_3}{\partial
r}~+~\frac{\partial^2 A_3}{\partial
z^2}~-~\frac{A_3}{r^2}\right)~=~A_3
\end{equation}
One can easily check that the two other components of $\Delta {\bf
A}$ are zero. The scalar equation  (\ref{eq:pep}) is either subject to
homogeneous Dirichlet boundary conditions or to the boundary constraint:
\begin{equation}\label{eq:bco}
\frac{\partial A_3}{\partial z}n_1~=~\left(\frac{\partial
A_3}{\partial r}~+~\frac{A_3}{r}\right)n_2
\end{equation}

The above equations can be simplified further by setting $B=rA_3$.
By this substitution our problem takes the
form:
\begin{equation}\label{eq:peb}
-\left(\frac{\partial^2 B}{\partial
r^2}~-~\frac{1}{r}\frac{\partial B}{\partial r}~+~\frac{\partial^2
B}{\partial z^2}\right)=B~~~{\rm in}~\Xi ~~~~~~~~~B=0 ~~{\rm on}~\partial\Xi
\end{equation}
where $\Xi$ is the section of the toroid region $\Omega$. Note
that $B$ is determined up to multiplicative constants (related to
the intensity of the final velocity field). On the left-hand side
of (\ref{eq:peb}) we have a positive definite operator, as a
result of little manipulation after multiplying by $B/r$ and
integrating on $\Xi$.
\par\smallskip

Since there is no dependence on $\phi$, we can simulate flows on
vortex rings having a symmetry axis coincident with the $z$-axis.
An analysis of this kind has been considered in \cite{funarol}
from a different perspective, where inviscid fluids are coupled
with electromagnetism. Note that the example introduced in section
1 for $n=1$ is related to famous Hertz solution, aimed to describe
the electromagnetic field emanated by an infinitesimal dipole.
Extensions, examined through a series of numerical tests, have
been successively investigated in \cite{chinosi}. The results and
the techniques there presented may be suitably adapted to the
study of vortex formation and stability.
\par\smallskip

For example, let $\Xi$ be the circle of radius $R$, situated in
the plane $(r,z)$ and centered at a point of the semi-axis $r>0$
very far away from the origin. By neglecting the terms in
(\ref{eq:pep}) containing $r$ at the denominator, the function
$A_3$ solves the 2D Laplace eigenvalue equation. The minimum
eigenvalue in $\Xi$ is approximately equal to $(\xi /R)^2$ (where
$\xi \approx 2.4$ is the first zero of $J_0$), so that, by taking
$R\approx 2.4$, we get a reasonably good solution of problem
(\ref{eq:rep}). Indeed, the 3D ring $\Omega$ with circular section $\Xi$ and
vertical axis $z$ is a domain where the equation $-\Delta {\bf
A}={\bf A}$ can be almost perfectly solved in terms of orthogonal
basis functions. There are infinite other rings, more or less of
the same shape and size as $\Omega$, where (\ref{eq:rep}) holds
once one knows the eigenfunctions of (\ref{eq:peb}). When the
section $\Xi$ approaches the origin, we can flatten the part of it
facing the $z$-axis, while preserving the minimum eigenvalue
$\lambda =1$ (see figure 2). The result is the transformation of
standard ring into a Hill's type one.
\par\smallskip

The flow in $\Omega$ may be defined by setting ${\bf v}=e^{-\nu
t}{\rm rot}{\bf A}$. Thus, as done in section 2, the full set of
Navier-Stokes equations (including the ${\rm div}{\bf v}=0$
condition) is satisfied by writing:
\begin{equation}\label{eq:na2}
\frac{\partial {\bf v}}{\partial t}~-~\nu \Delta {\bf v}~+~ ({\bf
v}\cdot\nabla ){\bf v}~=~{\textstyle{\frac{1}{2}}}\nabla\vert {\bf
v}\vert^2~-~{\bf v}\times {\rm rot}{\bf v}~=~-\nabla p~+~{\bf f}
\end{equation}
where the gradient term on the right-hand side has been
assimilated to a gradient of pressure, while the rest is a forcing
source. This last centripetal acceleration is the one needed to maintain
the flow constrained in the ring $\Omega$. The energy of the flow
is exponentially fading in time due to viscosity. In the above
mentioned procedure it is not clear how to prolong the field ${\bf
v}$ outside $\Omega$. Realistically, the external fluid is dragged
by viscous effects. The fluid across the boundary $\partial\Omega$
is continuous but may lose regularity there. This is the same
difficulty one may find in the study of typhoons, at the interface
between the central core and the peripheral dragged flow.
\par\smallskip

The request that the eigenvalue $\lambda$ remains equal to a given
constant is quite important in the framework of electromagnetic
applications (where $\lambda =c^2$ and $c$ is the speed of light),
but it is not crucial in fluid dynamics. We just have to pay
attention when passing to the parabolic equation $(\partial{\bf
v}/\partial t)=\nu \Delta {\bf v}$. For any fixed $t$, let ${\bf
w}$ be the stationary solution satisfying: $-\Delta {\bf
w}=\lambda (t){\bf w}$. As previously done, one sets ${\bf
v}=\gamma(t){\bf w}$. The function $\gamma$ has to be then
determined according to the relation:
\begin{equation}\label{eq:eql}
\frac{d \gamma}{dt}~=~-\nu\lambda\gamma ~~~~~{\rm for}~~t\in ]0,T[
\end{equation}
yielding the exponential $\gamma(t)=e^{-\nu t}$ in the
particular case $\lambda =1$.
\par\smallskip

\begin{center}
\begin{figure}[!h]
\centerline{\includegraphics[width=10.cm,height=8.cm]{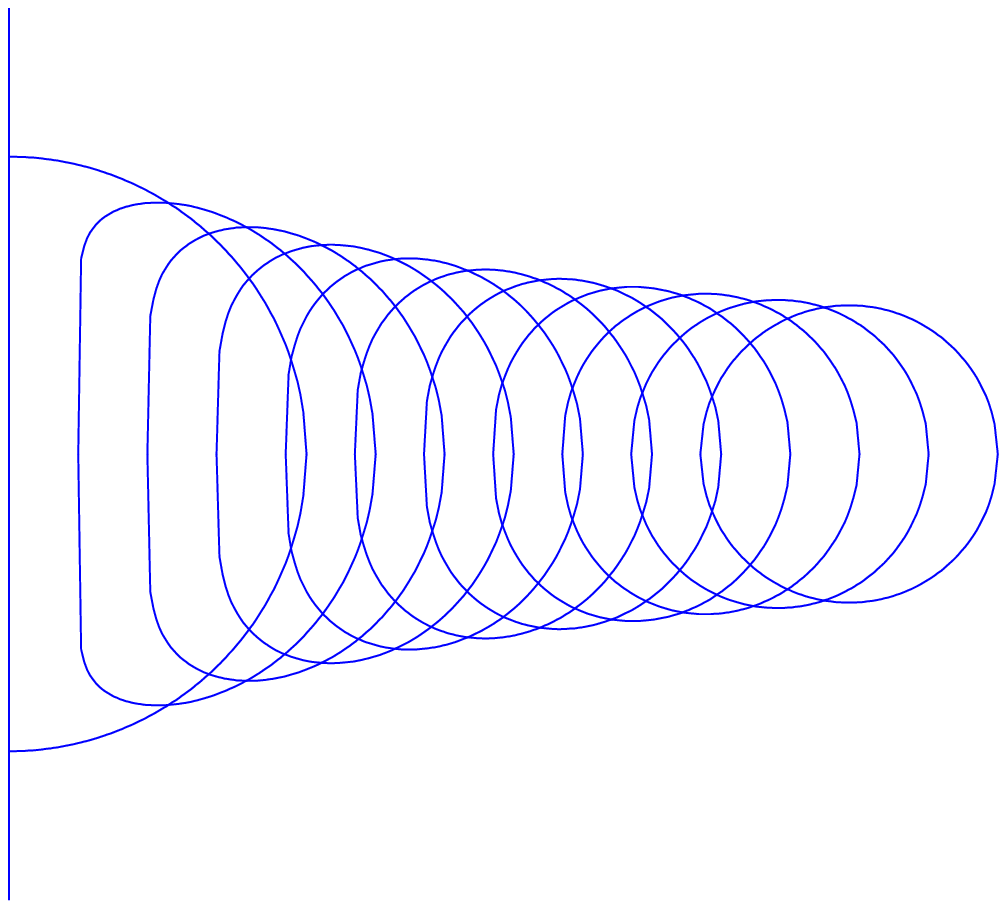}}
\vspace{-.3cm}
\begin{caption}{\small Sections of evolving tori. Starting from a classical ring
having circular section (right), one can deform the shape and
obtain through a sequence of intermediate rings a spherical
doughnut (left) where the hole is reduced to a vertical segment.}
\end{caption}
\end{figure}
\end{center}

A classical way of looking for troubled solutions of the
Navier-Stokes equations is to let evolve regular vortex tubes in
order to create a singularity at some point. Considerations about
this hypothesis are reported for instance in \cite{gala},
\cite{grauer}, \cite{kerr}, \cite{kerr2} and \cite{majda}, section
5.4. Some theoretical implications are examined in \cite{foxall}.
\par\smallskip

Therefore, from our viewpoint, one can imagine a smooth ring
driven by an internal force ${\bf f}_\tau$ changing with time. In
a finite time, the process ends up with the flow constricted into
a Hill's vortex, where the set of equations are actually solved on
a spherical domain. 
Note that, due to incompressibility, the volume of these rings
must remain constant. This is also true because we are imposing
tangential boundary conditions (see (\ref{eq:tbc})), therefore the
fluid cannot escape.
\par\smallskip

In the plane $(r,z)$,  a possible transition from a classical
vortex ring displaying circular section into a spherical vortex is
shown in figure 2. The evolution of the boundary may be described
for $t\in [0,1]$ by the formula:
\begin{equation}\label{eq:evo}
\left(\sqrt{(r-s(t))^2+z^2}\right)^{2-t}~=~(r-s(t))^{1-t}
\end{equation}
where $s>0$ is a time-dependent decreasing function satisfying
s(1)=0. For $t=0$, we have the circumference of radius
$\frac{1}{2}$ centered at $(\frac{1}{2}+s(0),0)$. After a sudden
change of topology, for $t=1$ we obtain the semi-circumference of
radius 1 centered at the origin. This construction is only given
to qualitatively illustrate the situation. We do not know for
example if the family of rings built by rotating the domains of
figure 2 around the $z$-axis preserves the eigenvalue $\lambda =1$
as prescribed by (\ref{eq:rep}).
\par\smallskip

It seems however that no irregularity turns out to be present at
the end of this process, since the solution presented in section 2
is smooth up to the origin. This does not prove that any attempt
to collapse a ring hole into a point (or a segment), according
to the equations, is going to fail. Nevertheless, the percentage
of success in the search of anomalous solutions of this type
becomes very low.

\section{A Hill's type vortex ring with twist}

We finish by retouching the solution defined in section 2. We now
add a component along the $\phi$ direction. Thus, we go back to
spherical coordinates $(r,\theta ,\phi )$ and recall the field
$\hat{\bf w}$ defined in (\ref{eq:vec}). For simplicity we only
discuss the case $n=1$. We now define:
\begin{equation}\label{eq:vem}
\hat{\bf w}~=~\left( \hat w_1,~\hat w_2,
~\sigma\frac{J_{3/2}(r)}{\sqrt{r}}\sin\theta\right)
\end{equation}
where $\sigma$ is a given number. Practically, the stationary
flow, while behaving as in figure 1, also moves around the
vertical axis. The third component in  (\ref{eq:vem}) is zero for
$r=0$ by virtue of (\ref{eq:as1}). By computing the rotational, we
now get:
\begin{equation}\label{eq:rom}
{\bf A}={\rm curl}\hat{\bf w}= \left( \frac{2\sigma J_{3/2}(r)}{r\sqrt{r}}\cos\theta ,
-\frac{\sigma}{\sqrt{r}}\Big( J^\prime_{3/2}(r)+\frac{J_{3/2}(r)}{2r}\Big)\sin\theta , ~A_3\right)
\end{equation}
where $A_3$ is given in (\ref{eq:poa}). It turns out that
$\hat{\bf w}~=~{\rm curl}{\bf A}$, so we are dealing with the same
kind of solutions previously studied. Such solutions are defined
everywhere. However, by denoting with $R$ the first zero of
$J_{3/2}$ (which is approximately equal to $4.49$), the solution
can be restricted to the sphere centered at the origin of radius
$R$.
\par\smallskip

Though $\hat{\bf w}$ still does not depend on $\phi$, the modified
flow field twists in the direction of the vector
$\boldsymbol{\phi}$ (see figure 3). Note that the boundary
conditions to be assigned on the surface of the sphere of radius
$R$ only contain the component along $\boldsymbol{\theta}$. As
figure 3 indicates, the field has only a vertical (along the
$z$-axis) component at the origin. In the immediate surroundings,
a component along $\boldsymbol{\phi}$ develops, which can be very
strong in magnitude since its strength depends on $\sigma$, which
is arbitrary. Therefore, sharp layers are observed near the
origin, although the solution continues to be smooth. Together
with the origin, other critical regions of sharp layer formation
are at the poles of the sphere.
\par\smallskip

The trajectory of a particle is practically untwisted near the
external boundary. In this situation the orbits are mildly precessing.
Nevertheless, as the particle gets to the pole, is pushed towards
the center and starts spiraling until it reemerges near the
opposite pole. For particles placed inside the sphere far from the
axis and the boundary, the orbits tend to follow a helicoid with
circular section, turning around the $z$-axis. There might be no
periodicity in this complex motion.
\par\smallskip

\begin{center}
\begin{figure}[!h]
\centerline{\includegraphics[width=12.cm,height=5.5cm]{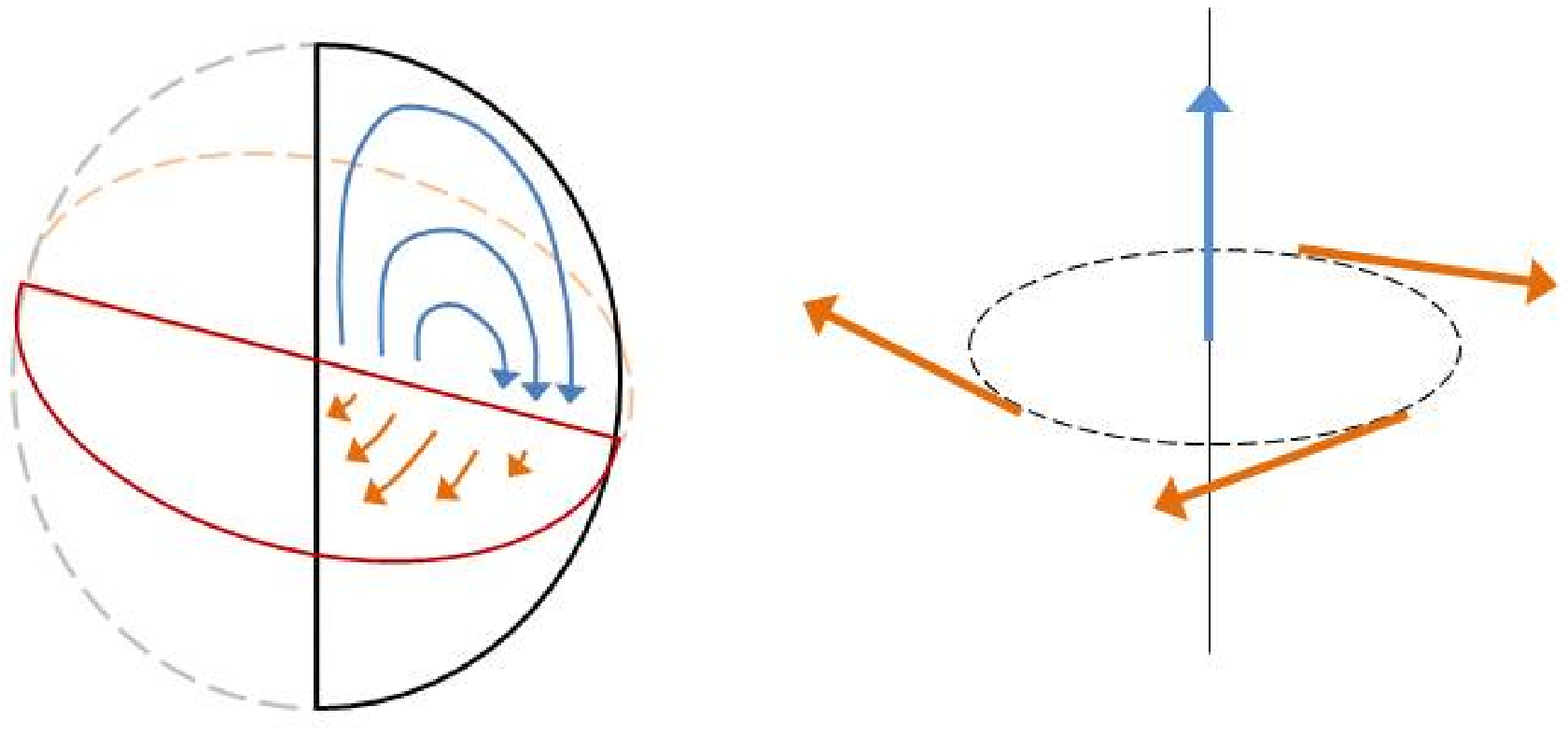}}
\vspace{-.1cm}
\begin{caption}{\small The field of the new spherical vortex combines the
component describing the rotation within each section with a
perpendicular one following closed circular patterns around the
vertical axis. The horizontal velocities are different depending
on the distance from the origin. In the neighborhood of the origin
(picture on the right), the trajectories pass with continuity from
a pure vertical motion to spirals. The variation can be
particularly significant, especially when $\sigma$ is large.}
\end{caption}
\end{figure}
\end{center}

When we compute the product $\hat{\bf w}\times {\rm rot}\hat{\bf w}$, it
should be evident that
its first two components tend to zero for $r\rightarrow 0$. The
third one takes the form:
\begin{equation}\label{eq:rom3}
(\hat{\bf w}\times {\rm rot}\hat{\bf w})_3~=~w_1A_2~-~w_2A_1
\end{equation}
A careful analysis shows that this also tends to zero for
$r\rightarrow 0$. We can argue locally in the neighborhood of the
origin by working in Cartesian coordinates. In this way, excluding
third-order terms, the
expression given in (\ref{eq:ort3}) has to be modified as follows:
\begin{equation}\label{eq:ort4}
\hat{\bf w}~
\approx~\frac{2}{3}\sqrt{\frac{2}{\pi}}\left(\frac{zx}{10}-\sigma  y,
~\frac{zy}{10}+\sigma x,~1-\frac{z^2}{10}-\frac{x^2+y^2}{5}\right)
\end{equation}
Considering that ${\rm rot}\hat{\bf w}\approx (2\sqrt{2}/3\sqrt{\pi})(-y/2, x/2, 2\sigma )$, one
ends up with the approximation:
$$-\nu\Delta \hat{\bf w}-\hat{\bf w}\times {\rm rot}\hat{\bf w}~\approx~
\frac{8}{9\pi}
\left( \frac{(4\sigma^2 -1)x}{2}+\frac{\sigma yz}{5}+\frac{xz^2}{20}+
\frac{x}{10}(x^2+y^2),\right.$$
\begin{equation}\label{eq:ort5}
~~~~~~~~\left. \frac{(4\sigma^2 -1)y}{2}-\frac{\sigma xz}{5}+\frac{yz^2}{20}+
\frac{y}{10}(x^2+y^2),~\frac{z}{20}(x^2+y^2)+\nu \frac{3}{2}\sqrt{\frac{\pi}{2}}\right)
\end{equation}
The special choice $\sigma =1/2$, that allows the above terms to
tend to zero faster at the origin,  worth to be recorded. For this
value of $\sigma$, the divergence of $\hat{\bf w}\times {\rm
rot}\hat{\bf w}$ is zero at the origin.
\par\smallskip

An interesting collateral problem to be studied is when the
initial solution of the homogeneous (${\bf f}=0$) Navier-Stokes
equations it taken to be equal to  $\hat{\bf w}$. In addition, the flow is
constantly supplied at the boundary of the sphere of radius $R$
with the stationary values of $\hat{\bf w}$. In absence of the
forcing term it is reasonable to expect a decay to zero at the
origin due to viscosity. This should be true independently of
$\sigma$,  but strange surprises could also happen. It would be
wise to run some numerical tests to verify the behavior.
\par\smallskip

Of course, one can add a twisting also to the annular flows
introduced in section 3. In cylindrical coordinates it is enough
to search for solutions of the form $\hat{\bf w}=(0,0,\hat w_3)$,
where $\hat w_3$ does not depend on $\phi$. If we require that
simultaneously ${\bf A}={\rm rot}\hat{\bf w}$ and $\hat{\bf
w}={\rm rot}{\bf A}$, we arrive at the equation:
\begin{equation}\label{eq:pew}
-\left(\frac{\partial^2 \hat w_3}{\partial
r^2}~+~\frac{1}{r}\frac{\partial \hat w_3}{\partial
r}~-~\frac{\hat w_3}{r^2}~+~\frac{\partial^2 \hat w_3}{\partial
z^2}\right)~=~\hat w_3
\end{equation}
to be solved in the section $\Xi$. The solution is determined up
to a constant $\sigma$. By imposing homogeneous Dirichlet boundary
conditions we get that the flow is swirling only at the interior
of the ring. For a ring with small circular section with respect
to the major diameter, the solution of the full Navier-Stokes
equations leads to a component along the $\phi$ direction similar
to that of a Poiseuille type flow. As before, the section $\Xi$
may be deformed depending on time as suggested by the plots of
figure 2, so ending up with a configuration as the one depicted in
figure 3.
\par\smallskip

In conclusion, we proposed a set of explicit solutions of the
non-homogeneous Navier-Stokes equations expressed in spherical
coordinates with the help of Bessel functions. These are defined
in the whole 3D space, but they can be easily restated in a
spherical domain subject to suitable tangential conditions. The
flow rotates forming a Hill's vortex remaining smooth up to the
inner point. A twist may be also introduced, but this does not
alter the regularity of the solution. This example suggests that
the search of blowing up solutions of the 3D Navier-Stokes
equations is not going to be easy, since most of the
counterexamples are built on the idea of crashing vortex tubes in
order to detect possible degenerations at some meeting point.
Together with the mentioned examples, we also proposed a way to
explicitly construct vortex rings, that can be helpful in various
other circumstances, both for theoretical and computational
aspects.

\end{document}